# Extremely High Thermal Conductivity of Aligned Bulk Carbon Nanotube-Polyethylene Composites

Quanwen Liao[1], Zhichun Liu[1], Wei Liu[1], Chengcheng Deng[1], and Nuo Yang[1,2]

1 School of Energy and Power Engineering, Huazhong University of Science and Technology (HUST), Wuhan 430074, China

2 State Key Laboratory of Coal Combustion, Huazhong University of Science and Technology (HUST), Wuhan 430074, China

a) Electronic mail: nuo@hust.edu.cn

b) Electronic mail: zcliu@hust.edu.cn

**ABSTRACT**

The poor thermal conductivity of bulk polymers may be enhanced by combining them with high thermal conductivity materials such as carbon nanotubes. Different from random doping, we find that the aligned carbon nanotube-polyethylene composites (ACPCs) has a high thermal conductivity by non-equilibrium molecular dynamics simulations. The analyses indicate that the aligned structure can not only take advantage of the high thermal conduction of carbon nanotubes, but enhance thermal conduction of polyethylene (PE) chains. Our predictions may inspire manufacturing aligned polymer-based composites for a wide variety of applications.

**Introduction**

Polymers have been widely utilized in all walks of life owing to their outstanding physical properties, such as high toughness, low density, and corrosion resistance. However, its ultra-low thermal conductivities (κ), on the order of 0.1 Wm$^{-1}$K$^{-1}$ at room temperature[1], limits its applications.[2, 3] Because the random orientation and mutual twining of polymer structures lead to much more phonon scatterings and a short phonon mean free path.

Recently, it is reported that a suspended polymer chain and oriented polymer chains have remarkable thermal conductivities.[4-7] Chen's group predicted that the κ of a suspended polyethylene chain (SPEC) achieved as high as 350 Wm$^{-1}$K$^{-1}$ at room temperature by molecular dynamics (MD) simulation[4, 5] and measured the κ of ultra-drawn polyethylene (PE) nanofibers as 104 Wm$^{-1}$K$^{-1}$ using the cantilever method[6]. Virendra *et al.* measured the κ of amorphous polythiophene nanofibers with room-temperature as 4.4 Wm$^{-1}$K$^{-1}$ and calculated a suspended polythiophene's κ as 43.3 Wm$^{-1}$K$^{-1}$ by MD.[7] Moreover, Zhang *et al.* demonstrated that a high thermal conductivity and good stability could be achieved in polymers with rigid backbones.[3] Although the polymers' thermal conductivity is enhanced, it is quite difficult to take advantage of the suspended polymer chain or oriented polymer chains in realistic applications.

Another way to obtain polymer structures with enhanced thermal properties is the polymer/carbon nanotube nanocomposites. The carbon nanotube (CNT), with a super-high thermal conductivity on the order of 1000 Wm$^{-1}$K$^{-1}$ at room temperature,[8] has been well studied since its discovery in 1991.[9,10] Some efforts have been made in fabricating polymer/CNT nanocomposites which could combine the high thermal transport properties of CNTs with the excellent mechanical properties of polymers.[11-23] The interfacial thermal resistance between the CNT and surrounding polymer matrix plays a crucial role in obstructing the thermal transport in nanocomposites.[21, 24, 25] It was suggested that a strong coupling between CNTs or other additives and polymers could reduce phonon scattering at interfaces, and effectively improve the κ of nanocomposites.[24, 25]

In order to enhance thermal properties of PE composites, we investigated numerically the thermal conductivity of aligned carbon nanotube-polyethylene composites (ACPCs) in this paper. Here, the well-studied (10, 10) single wall carbon nanotube (SWNT) and PE chains are chosen. In ACPC models, the SWNTs are aligned with PE chains, which could avoid both the disordered and interfacial scatterings of phonon in the amorphous PE matrix. Due to the reinforcement from SWNTs, the ACPC structures could vibrate like a phonon crystal and have a high thermal property.

In the following, we provide firstly a description of the model and simulation procedures. Secondly, we show the simulation results and analyzing mechanism. Our results show that the non-bonded interactions between the parallel-aligned SWNT and PE chains could enhance significantly the thermal conductivity of PE chains. Our study may inspire productions and measurements of aligned carbon nanotube and polymer-based composites.

**Methods**

Classical non-equilibrium molecular dynamics (NEMD) simulations are used to study the thermal conductivity of the SWNT, the SPEC and the ACPC. All simulations are performed by the large-scale atomic/molecular massively parallel simulator (LAMMPS) package developed by Sandia National Laboratories.[26] The temperatures of the heat source and heat sink are set at 310 K and 290 K, respectively. The fixed boundary condition is applied in the longitudinal direction and the periodic boundary conditions are applied in the two transversal directions. The cross-section of ACPC simulation cell after relaxation spreads from 400 to 500 Å$^2$ which depends on the structure of ACPC.

The potential energy of the SWNT is described by a Morse bond and a harmonic cosine angle for bonding interactions, which includes both the two-body and three-body potential terms.[27-29] The atomic interactions of PE chains are described by an adaptive intermolecular reactive empirical bond order (AIREBO) potential,[30] which is developed from the second-generation Brenner potential.[31] In addition, the non-bonded interactions between the SWNT and PE are described by the Lennard-Jones potential:

$$V_{LJ}(r_{ij}) = 4\varepsilon[(\sigma/r_{ij})^{12} - (\sigma/r_{ij})^6] \qquad (1)$$

Where $\varepsilon$ is the depth of the potential well, $r_{ij}$ is the distance between atom i and j. The Lennard-Jones parameters are $\sigma_{SC-PC}$ = 3.4 Å, $\varepsilon_{SC-PC}$ = 0.0028 kcal/mol, $\sigma_{SC-PH}$ = 3.025 Å, and $\varepsilon_{SC-PH}$ = 0.0021 kcal/mol; the SC, PC, and PH subscripts represent the carbon atoms within the SWNT, the carbon atoms within the PE chains, and the hydrogen atoms within the PE chains, respectively. Additionally, an 8.5 Å cutoff distance is used for the 12-6 Lennard-Jones interaction.

Fig. 1 (a) shows the typically perspective view of initial positions of ACPC 3-8. After the relaxation, the final structures of ACPC 1-12 and 3-12 are shown in Fig. 1 (b) and (c), respectively. Moreover, several different ACPC structures are taken into consideration. We named ACPC M-N as the structure which have M PE chains inside the SWNT and N chains outside the SWNT. Fig. 1 (d) shows a typical setup and the corresponding temperature profile. The simulation system is divided into 20 or 50 slabs according to the length.[32] The motion equations are integrated by the Velocity Verlet algorithm with a time step of 0.2 fs. The NEMD method has been detailed in the Ref. 33.

In calculations of thermal conductivity, it is based on Fourier's law, $\kappa = -J/(A \cdot \nabla T)$, where J is heat flux, A is cross-section area, and T is temperature. The cross-section of (10, 10) SWNT is defined as a ring with 3.4 Å in thick.[34] Besides, the cross-sectional area of a PE chain is taken as 18 Å$^2$.[4]

**Results and discussions**

The main results are shown in Fig. 2 which includes the thermal conductivities of a suspended SWNT, a SPEC, and several ACPCs with different structures. The value of $\kappa_{SWNT}$ reaches 155 Wm$^{-1}$K$^{-1}$ when the length of the SWNT is 160 nm. Obviously, the κ of (10, 10) SWNT is not converged and will keep diverging as its length increases. Our result is slightly smaller than previous reported simulation results,[35-38] due to the difference of empirical potential. The simplified Morse potential neglects some interactions within the SWNT, such as dihedral and van der Waals interactions. That is, our result is conservative and undervalue the $\kappa_{SWNT}$ a little.

Similarly, the thermal conductivity of a SPEC also shows a strong length dependence. The $\kappa_{SPEC}$ achieves 57 Wm$^{-1}$K$^{-1}$ with a length of 160 nm at room temperature. Compared with previous simulation results, our result is less than the Hu's[2], 104 Wm$^{-1}$K$^{-1}$ at 160 nm length, and a little higher than the Zhang's[39], 49 Wm$^{-1}$K$^{-1}$ with at 50 nm length. The discrepancy between them is chiefly derived from the different models used for the PE chain. Such as, a simplified model of a PE chain is applied in Hu's simulations, where methylene ($CH_2$) groups are regarded as united atoms. Moreover, in Zhang's work, a different potential (COMPASS) is used to model the PE chain. As a SPEC possesses a much higher thermal conductivity than that of an amorphous bulk PE, we will take advantage of this property in enhancing the κ of PE-based materials.

The most significant finding is that the thermal conductivities of ACPCs are not only three orders higher than the bulk PE, but almost twice as large as a SPEC. That is, the κ of PE composite is greatly enhanced by the SWNT's reinforcement. In our simulations, the maximum value of $\kappa_{ACPC}$ is 99.5 Wm$^{-1}$K$^{-1}$ for an ACPC 3-8 with a length of 320 nm, which is comparable to that of measurements in ultra-drawn PE nanofibers, around 104 Wm$^{-1}$K$^{-1}$.[6] Moreover, the $\kappa_{ACPC}$ is just limited by the simulation cell's length and could reach a much more higher value with the increasing of length due to the divergence behavior of κ in low dimensional structures.[40] Furthermore, there are few reports on polymer nanocomposites with such a high thermal conductivity and the $\kappa_{ACPC}$ is at least 30 times higher than the reported κ of CNT-polymer nanocomposites.[20, 22, 23]

The high thermal conductivity of ACPC attributes to three mechanisms. Firstly, the SWNTs possess a high thermal conductivity, which contributes a lot in enhancing the thermal conductivity of the PE-based nanocomposites. Secondly, instead of random doping, the SWNTs are aligned with PE chains, which is the most important factor. The aligned structures not only take advantage of the divergent κ of PE chains with length, but avoid the interface scattering issue between SWNTs and PE chains in nanocomposites. Thirdly, it was found that the non-bonded interactions between the SWNTs and PE chains also have a significantly positive effect on the thermal transport in ACPCs. The van der Waals forces between the SWNTs and PE chains hinder vibrations, inducing a crystal-like structure in the PE chains. Hence, the thermal conductivity of the PE chains within an ACPC is improved by the SWNT interactions to become even higher than that of a SPEC.

Besides the length dependence, the thermal conductivity of ACPC also depends on the number of chains inside SWNTs, M (shown in Fig. 2). As the number M increases, the $\kappa_{ACPC}$ first increases and then decreases. A maximum value of thermal conductivity was observed when there are three PE chains inside SWNTs. Due to the space limitation inside a SWNT, the van der Waals interactions increase with an increasing number of PE chains within a SWNT. The van der Waals interactions could take two competitive effects. When 3 or fewer chains are placed inside the SWNT, there is a slight van der Waals interaction which can suppress the transversal bending of chains and enhance the heat transfer. However, when M is above 3, stronger interactions will bring more phonon scatterings which decrease the thermal conductivity.

In the following, we show a further analysis of the mechanism in the thermal conductivity enhancement of PE chains within an ACPC. As shown in superimposed images (inserts of Fig. 3), the PE chains within ACPC 3-12 have a clear crystalline structure compared with the SPEC. Accordingly, shown in Fig. 3, the radius density profile, g(r), of a SPEC appears amorphous, suggesting a large spread of atom vibrations and many segmental rotations of chain. In contrast, the g(r) of a PE chain within ACPC 3-12 has clear peaks and valleys, corresponding to a more ordered crystal lattice. That is, the van der Waals forces in ACPC make PE chains crystal-like and reduce disorder phonon scatterings survived in suspended chain.

We keep analyzing the details of the enhancement of thermal conductivity by ACPCs. In Fig. 4 (a), it shows that the thermal conductivity of 20 nm ACPC 3-N changes a little on No. of PE chains outside the SWNT (N). A maximum value of 63.7 Wm$^{-1}$K$^{-1}$ was obtained for the ACPC 3-N thermal conductivity when N is 8. Besides, we pick up the heat flux (J) of PE chains alone in ACPCs. The thermal conductance is defined as $G = \kappa \cdot A / L = -J / \Delta T$, where $\kappa$, $A$, and $L$ are the thermal conductivity, cross-section area, and length, respectively. The thermal conductance of PE chains in ACPCs (the blue circles) is compared with that of a 20 nm SPEC (blue dashed line) shown in Fig. 4 (a). It shows that there is a significant enhancement in the thermal conductance of PE chains in the ACPCs. The $G_{PE}$ of ACPC 3-4 is around four times larger than G of SPEC, since the non-bonding interactions in ACPC make a more crystal-like PE structure. That is, the high $\kappa$ of ACPC comes from not only SWNT but the PE chains.

Fig. 4 (b) shows the thermal conductivity of ACPCs versus PE content for ACPC M-12 and 3-N structures. With increasing PE content, the thermal conductivity of ACPCs does not decrease monotonically, although $\kappa_{PE}$ is much smaller than $\kappa_{SWNT}$. For example, the thermal conductivity of 20 nm ACPC 3-N is not sensitive to the increase of PE content. For the 20 nm and 40nm ACPC M-12, an increase is observed when the PE content increase from 19.5% to 22%. Therefore, the PE chains in ACPCs do account for a significant contribution in thermal transport.

Moreover, Fig. 4 (c) shows the contribution of PE chains and SWNT to thermal conductance of ACPC 3-8 and ACPC 3-4 with 80 nm in length. It shows that the PE chains contribute a considerable percentage of the total thermal transport, 36.4% (27.8%) for ACPC 3-8 (3-4). Both the G values of SWNTs in ACPC 3-4 (22.6 × $10^{-10}$ WK$^{-1}$) and ACPC 3-8 (23.86 × $10^{-10}$ WK$^{-1}$) are smaller than that of a suspended 80 nm SWNT (26.32 × $10^{-10}$ WK$^{-1}$) due to the scattering from non-bonding interactions in ACPC. The values of thermal conductance per PE chain inside and outside the SWNT in ACPC 3-8 and 3-4 are shown in Fig. 4 (d). Compared with an 80nm SPEC, the G value per PE chain is improved by as large as 38.5% for chains inside SWNT in ACPC 3-8. That is, the non-bonded interactions in ACPCs enhance the thermal transport of the PE considerably, about 23% on average.

**Conclusions**

The thermal conductivity (κ) of aligned carbon nanotube-polyethylene composites are studied by non-equilibrium molecular dynamics simulations. The most significant finding is that the thermal conductivities of ACPCs are not only three orders higher than the bulk PE, but almost twice as large as a suspended PE chain. The $\kappa_{ACPC}$ is also at least 30 times higher than the κ of other CNT-polymer nanocomposites. Besides, there is a large enhancement (~23%) of thermal conduction for PE chains in ACPCs even comparing with a suspended PE chain which is well known by its high κ. So that, the PE chains have a considerable contribution (~30%) to the thermal transport in ACPCs.

The high thermal conductivity of ACPCs attributes to the high thermal conductivity SWNTs, the aligned SWNTs with PE chains instead of random doping, and the non-bonded interactions between SWNTs and PE chains. Our predictions may inspire manufacturing aligned polymer-based composites for a wide variety of applications.

**Acknowledgments**

Z. L. was supported in part by the National Natural Science Foundation of China (Grant No. 51376069) and by the Major State Basic Research Development Program of China (Grant No. 2013CB228302). N. Y. was supported in part by the National Natural Science Foundation of China (Grant No. 11204216) and the Self-Innovation Foundation of HUST (Grant No. 2014TS115). The work was performed at the National Supercomputer Center in Tianjin and the calculations were performed on TianHe-1(A).

**Figure Captions**

Fig. 1 Schematic view of the aligned carbon nanotube-polyethylene composites (ACPCs). (a) The perspective view of structure of ACPC 3-8; (b) and (c) The orthographic views of the relaxed structures of ACPC 1-12 and 3-12, respectively; (d) The temperature profile of ACPC by non-equilibrium molecular dynamics (NEMD).

Fig. 2 The thermal conductivities ($\kappa$) of a (10, 10) single wall carbon nanotube (SWNT), a suspended polyethylene chain (SPEC), and ACPCs versus the simulation cell's length. We named ACPC M-N as the structure which have M polyethylene (PE) chains inside the SWNT and N chains outside the SWNT. The thermal conductivities of an ACPC M-N are between those of the SWNT and SPEC. The ACPC 3-8 and ACPC 3-12 have higher thermal conductivities comparing with other ACPC structures. The $\kappa_{ACPCs}$ are not only three orders higher than the bulk PE, but almost twice as large as a SPEC.

Fig. 3 The radial atomic density profiles, g(r), for a SPEC and a PE chain within ACPC 3-12. The red (grey/green) scatters are the superimposed atoms image for a 10 nm SPEC (ACPC 3-12), respectively. The atoms images are established by stacking fifteen frames in a simulation over time.

Fig. 4 (a) The thermal conductivity ($\kappa$) of ACPC 3-N versus the numbers of chains outside the SWNT (N). The thermal conductance (G) per PE chain compared to a SPEC at 20 nm length; (b) The dependence of $\kappa_{ACPC}$ on PE content for four different ACPC structures; (c) The contributions of SWNT and PE to the total thermal conductance in an 80 nm length ACPC 3-4 (ACPC 3-8). The PE accounts for 27.8% (ACPC 3-4) and 36.4% (ACPC 3-8), respectively; (d) The thermal conductance per PE chain for chains inside (outside) SWNT in an 80 nm length ACPC 3-4 (ACPC 3-8). The blue dashed line corresponds to the G of an 80 nm SPEC.

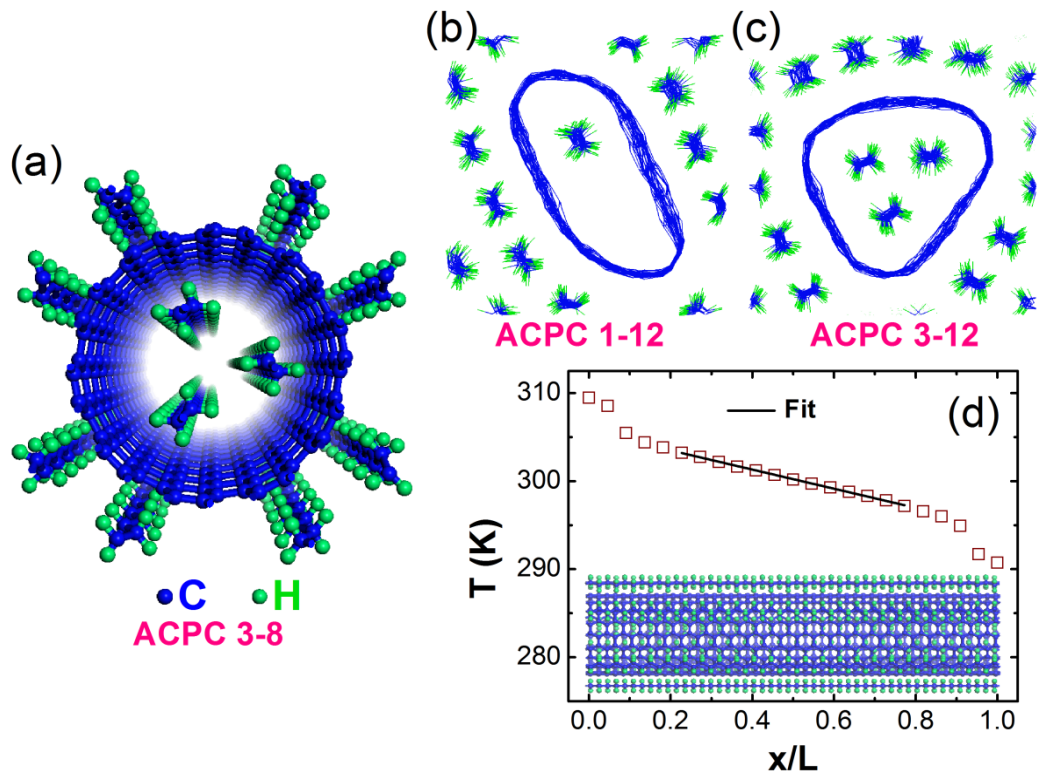

**Fig. 1**

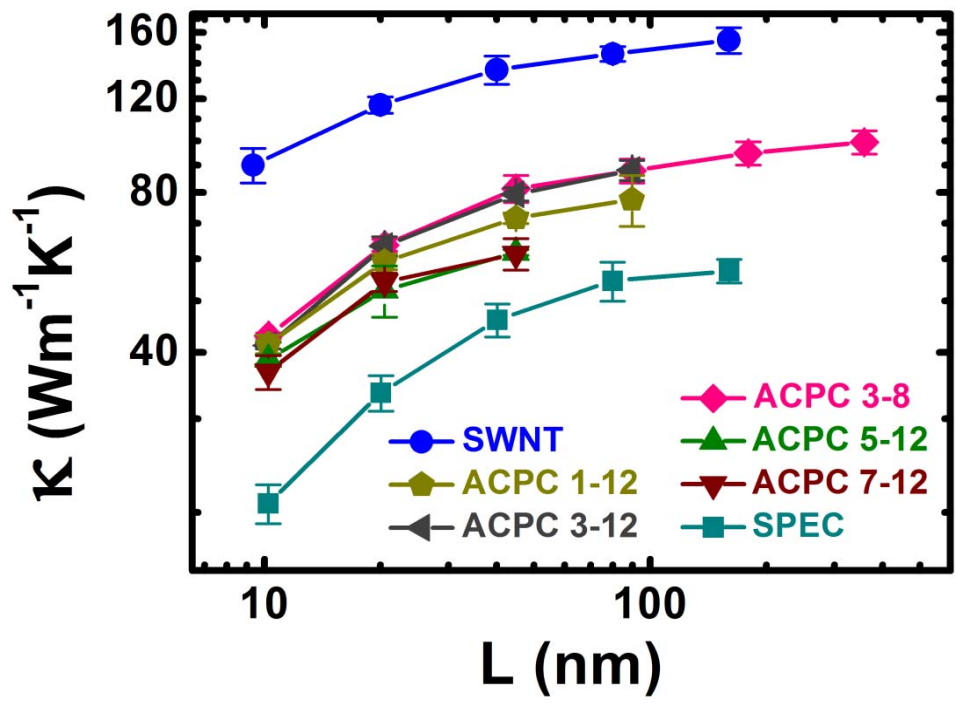

**Fig. 2**

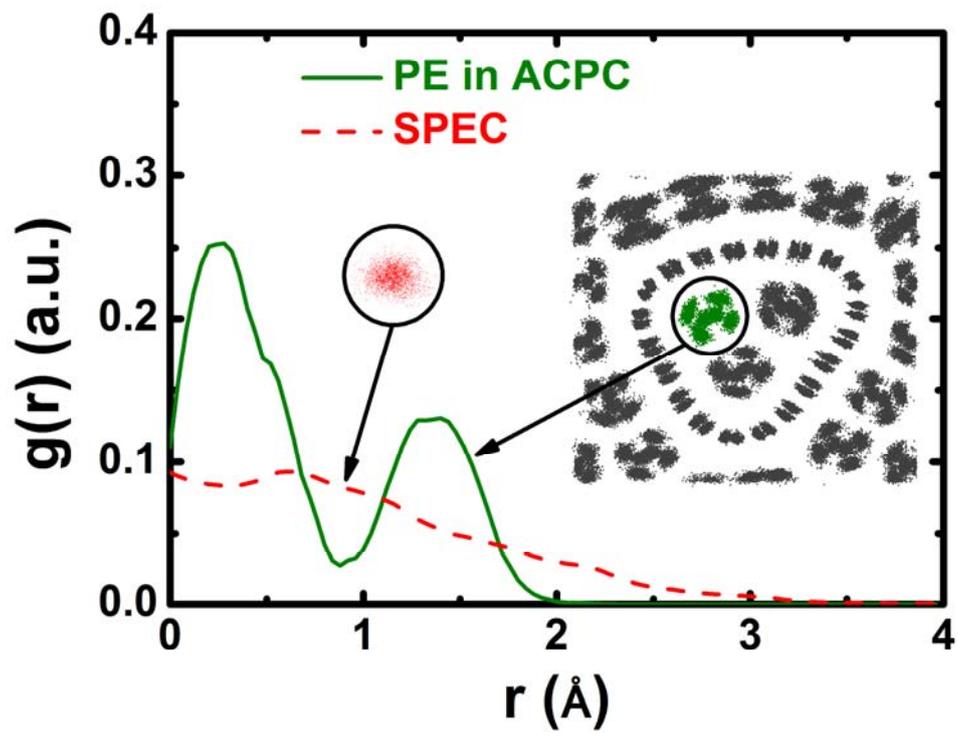

**Fig. 3**

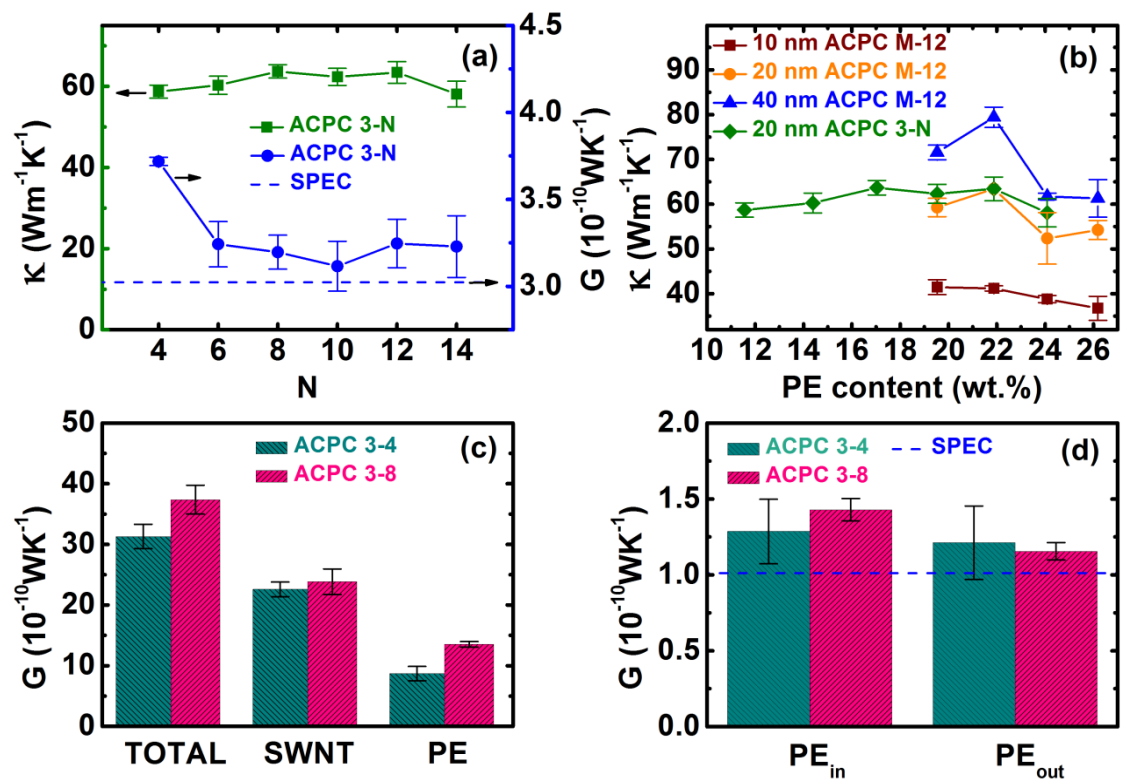

**Fig. 4**